\documentclass[entropy,article,accept,pdftex,moreauthors]{Definitions/mdpi} 
\firstpage{1} 
\pubvolume{1}
\issuenum{1}
\articlenumber{0}
\pubyear{2025}
\copyrightyear{2025}
\datereceived{ } 
\daterevised{ } 
\dateaccepted{ } 
\datepublished{ } 
\hreflink{https://doi.org/} 

\Title{Dual effects of Lamb Shift in Quantum Thermodynamical Systems}

\TitleCitation{Title}

\Author{Zi-chen Zhang $^{1}$\orcidA{} and Chang-shui Yu $^{1,}$*}

\address{$^{1}$ \quad $^1$School of Physics, Dalian University of Technology, Dalian 116024, P.R. China; 1278493906@mail.dlut.edu.cn}

\corres{Correspondence: ycs@dlut.edu.cn}

\abstract{ The Lamb shift as an additional energy correction induced by environments usually has a marginal contribution and hence is neglected. We demonstrate that the Lamb shift, which modifies the energy levels, can influence the heat current to varying extents. We focus on the steady-state heat current through two coupled two-level atoms, respectively, in contact with a heat reservoir at a certain temperature.  We find that the Lamb shift suppresses the steady-state heat current at small temperature gradients, while at large gradients, the heat current is restricted by an upper bound without the Lamb shift but diverges when it is included. These results not only demonstrate the Lamb shift’s critical role in quantum heat transport but also advance our understanding of its impact in quantum thermodynamics.}

\keyword{Lamb shift; heat current; quantum thermodynamics } 

\begin{document}


\section{Introduction}
In realistic scenarios, the physical system can hardly be completely isolated from its external environment. It is inevitable for an open system to exchange energy, information, or matter with its surroundings.  The evolution of open systems is determined not only by the system's intrinsic Hamiltonian but also by its coupling with environmental degrees of freedom \cite{breuer2002theory}. There are a variety of nontrivial quantum phenomena, including decoherence and energy dissipation, in open systems. The master equation (ME) is one of the critical methods describing the dynamic evolution of open systems \cite{breuer2002theory,rivas2010markovian,carmichael2013statistical}. The most popular ME could count on 
the Gorini-Kossakowski-Lindblad-Sudarshan (GKLS) ME \cite{REDFIELD19651,lindblad1976generators,gorini1976completely} based on the Born-Markov-Secular approximation, which
has been extensively validated and applied in numerous studies \cite{gemmer2009quantum,alicki2006internal,PhysRevLett.122.150603,albash2012quantum,10.1063/5.0259168,PhysRevE.107.014108}. Some other approaches, such as the coarse-graining
method \cite{mozgunov2020completely}, the universal Lindblad equation 
\cite{PhysRevB.102.115109}, and the Geometric-arithmetic master equation \cite{Davidović_2022, Davidovic2020completelypositive} are also used to derive the master equation
without the secular approximation. In addition, some works especially  focused on examining the validity of the ME \cite{lidar2019lecture,mozgunov2020completely,Davidovic2020completelypositive,PhysRevB.102.115109,PhysRevA.105.032208,PhysRevA.101.012103}. Among these contributions, one can usually find the system's energy level shift, known as the Lamb shift, caused by its interaction with its environment. 

The Lamb shift typically manifests as an additional term as the energy level correction caused by the environment, which usually has a marginal contributions on the question of interest \cite{breuer2002theory,rivas2010markovian,carmichael2013statistical}. A numerical demonstration showed that the Lamb shift only weakly perturbs the system Hamiltonian \cite{PhysRevB.102.115109}; Refs. \cite{PhysRevE.107.014108, PhysRevA.105.022424}  ignored the contribution of the Lamb shift by comparing the order of magnitude of the Lamb shift in the weak coupling limit; Some studies have assumed that heat baths possess a huge bandwidth, resulting in an effective zero Lamb shift \cite{PhysRevA.103.052209}. In this sense, the Lamb shift is safely neglected in many applications. For example, the ME, which neglects the Lamb shift, is used to analyze the interactions between light and matter, quantum interference of light, and quantum light sources, such as single-photon sources and lasers \cite{PhysRevLett.109.193602, PhysRevLett.110.243601, PhysRevLett.106.013601}. The decoherence in various quantum information processes \cite{PhysRevLett.95.250503, PhysRevA.65.012322, PRXQuantum.4.040329, PhysRevA.108.053711} has been analyzed without considering the Lamb shift. The ME without the Lamb shift is used to study the energy exchange processes between the working substance and its heat reservoirs \cite{PhysRevLett.116.200601, PhysRevLett.120.200603, PhysRevLett.127.190604,guo2018quantum, PhysRevA.98.052123, PhysRevE.95.022128, PhysRevA.107.032602}.  The Lamb shift commutes with the system's Hamiltonian; hence, the steady state is independent of the Lamb shift, and so is the steady-state heat current \cite{PhysRevApplied.16.034026}. Recently, quantum batteries have attracted increasing interest \cite{PhysRevA.102.052223, PhysRevB.99.035421, PhysRevA.103.033715,morrone2023charging,6c73-ll23}, where ME facilitates understanding the charging and discharging processes of quantum states, especially considering the environmental factors such as temperature and noise on battery performance, but one can find that the Lamb shift isn't considered either.

 However, recent studies have begun to reevaluate the significance of the Lamb shift, acknowledging its crucial role in accurately characterizing system–environment interactions. Numerical evidence has shown the substantial impact of the Lamb shift \cite{albash2012quantum}. For instance, under specific parameter choices, the refined Lamb shift Hamiltonian yields more accurate results than models neglecting this contribution \cite{PhysRevA.103.062226}. The discrepancies observed between Born–Markov methods and the stochastic Liouville equation with dissipation (SLED) at low temperatures have been attributed to the omission of the Lamb shift \cite{PhysRevB.103.214308}. Moreover, a significant collective Lamb shift was experimentally demonstrated using two distant superconducting qubits \cite{PhysRevLett.123.233602}. In systems such as a giant artificial atom with multiple coupling points, the Lamb shift becomes a pronounced, frequency-dependent, and engineerable quantity that actively influences relaxation rates and enables tunable anharmonicity \cite{PhysRevA.90.013837}. These energy shifts play an essential role in quantum thermodynamics. For example, in electric circuit systems \cite{doi:10.1126/science.1164482, PhysRevA.84.052103}, quantum refrigeration \cite{PhysRevLett.105.130401, yu2014re}, quantum dots \cite{thierschmann2015three}, and giant atoms \cite{kannan2020waveguide, PhysRevA.90.013837}, accurate accounting of the Lamb shift is critical for predicting heat currents and other thermodynamic quantities reliably \cite{23nd-hp2n}.

In this paper, we find the significant influence of the Lamb shifts on the
heat transport through two coupled atoms interacting with a heat bath
respectively. Here, we do not distinguish between the Lamb and Stark shifts in this paper for simplicity, but rather collectively refer to the total of the environment-induced frequency shift of the system as the Lamb shift.  We find that the heat current approaches an
upper bound with the temperature difference increasing when the Lamb shift
isn't considered as usual \cite{quiroga2007nonequilibrium,sinaysky2008dynamics,hu2018steady,yang2022heat}.
In contrast, the heat current will monotonically increase with the
temperature difference increasing if we consider the Lamb shift. This
difference in heat currents persists when other forms of spectral densities are taken. 

This paper is organized as follows. In Sec. \ref{2}, we give a brief
description of our model and derive the master equations under the
Born-Markov-Secular approximation. In Sec. \ref{3}, we calculate the Lamb shift. In Sec. \ref{4}, we compare the heat currents with and without the Lamb shift. We conclude with a summary in Sec. \ref {6}.

\section{The Model}
\label{2}
\begin{figure}[t]
\centering\includegraphics[width=9cm]{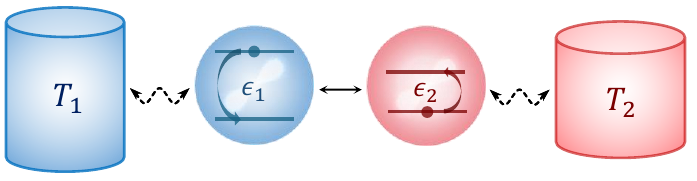}
\caption{The schematic illustration of our model, where the dashed line
represents weak coupling, and the solid line represents strong coupling. The
temperatures of two heat reservoirs are $T_{1}$ and $T_{2}$, the energy
separation of two qubits are $\protect\varepsilon _{1}$ and $\protect%
\varepsilon _{2}$.}
\label{example}
\end{figure}

Let's consider two coupled two-level atoms (TLAs) interacting with a distinct thermal reservoir, respectively, as shown in Fig. \ref{example}.  The model has been widely studied in various cases \cite{PhysRevLett.100.105901, PhysRevLett.88.094302, PhysRevE.83.031106, PhysRevB.100.045418}. The total Hamiltonian of the system and the reservoirs is 
\begin{equation}
{H}={H}_{S}+{H}_{B1}+{H}_{B2}+{H}_{SB1}+{H}_{SB2},
\end{equation}
where 
\begin{equation}
{H_{S}}=\frac{{\varepsilon _{1}}}{2}\sigma _{1}^{z}+\frac{{\varepsilon _{2}}%
}{2}\sigma _{2}^{z}+g \sigma _{1}^{x}\sigma _{2}^{x},
\end{equation}
with $\varepsilon _{1}\geq \varepsilon _{2}$, $\sigma _{i}^{z}$ and $\sigma _{i}^{x}$ denoting the Pauli matrices, and $g 
$ denoting the coupling strength of two qubits. For simplicity, we adopt natural units by setting the reduced Planck constant $\hbar = 1$ and the Boltzmann constant $k_{B} = 1$. 
${H}_{B1}$ and ${H}_{B2}$ are the Hamiltonians of the two thermal reservoirs, which is characterized by a collection of independent harmonic oscillators, where \begin{equation}
{H}_{Bj}=\sum_{n}\omega _{n,j}{b}_{j,n}^{\dagger }{b}_{j,n}
\end{equation}
with the summation running over all discrete modes of the $j$th thermal reservoir. ${H}_{SB1}$ and ${H}_{SB2}$ are the interaction Hamiltonians between the system of interest and the reservoirs, where 
\begin{align}
{H_{S{B_{j}}}}=\sigma _{j}^{x}\sum\limits_{n}{{g_{j,n}}\left( {{b_{j,n}}%
+b_{j,n}^{\dag }}\right) }=\sigma _{j}^{x}B_{j}^{x}
\end{align}%
with $B_{j}^{x}=\sum\limits_{n}{{g_{j,n}}\left( {{b_{j,n}}+b_{j,n}^{\dag }}%
\right) }$.

To obtain the dynamics of the system, we'd like to derive the master
equation. Following the standard process \cite%
{breuer2002theory,schaller2014non}, we would like to first give the eigensystems of $H_S$ as
$H_S\left\vert s_i\right\rangle=s_i\left\vert s_i\right\rangle$, where the eigenvectors read
\begin{eqnarray}
  \left| {{s_1}} \right\rangle  = \cos \frac{\varphi }{2}\left| {0,0} \right\rangle  - \sin \frac{\varphi }{2}\left| {1,1} \right\rangle ,\left| {{s_2}} \right\rangle  = \sin \frac{\varphi }{2}\left| {0,0} \right\rangle  + \cos \frac{\varphi }{2}\left| {1,1} \right\rangle , \hfill \\
  \left| {{s_3}} \right\rangle  = \cos \frac{\theta }{2}\left| {1,0} \right\rangle  + \sin \frac{\theta }{2}\left| {0,1} \right\rangle ,\left| {{s_4}} \right\rangle  =  - \sin \frac{\theta }{2}\left| {1,0} \right\rangle  + \cos \frac{\theta }{2}\left| {0,1} \right\rangle , \hfill \\ 
\end{eqnarray}
and the eigenvalues are ${s_{1}}=-\beta ,{s_{2}}=\beta ,{s_{3}}=\alpha ,{s_{4}}=-\alpha $ with 
\begin{equation}
\alpha =\sqrt{\frac{{{{\left( {{\varepsilon _{1}}-{\varepsilon _{2}}}\right) 
}^{2}}}}{4}+{g ^{2}}},\beta =\sqrt{\frac{{{{\left( {{\varepsilon _{1}}+{%
\varepsilon _{2}}}\right) }^{2}}}}{4}+{g ^{2}}},
\tan \varphi =\frac{{2g }}{{{\varepsilon _{1}}+{\varepsilon _{2}}}},\tan
\theta =\frac{{2g }}{{{\varepsilon _{1}}-{\varepsilon _{2}}}}.
\end{equation}
Thus, we can derive the eigenoperators 
${V}_{j\mu }$ with  $[{H}_{S},{V}_{j\mu}]=-\omega _{j\mu }{V}_{j\mu }$ \cite{breuer2002theory,schaller2014non} as 
\[\begin{gathered}
  {V_{1,1}} = \sin {\phi _ + }\left( {\left| {{s_3}} \right\rangle \left\langle {{s_2}} \right| - \left| {{s_1}} \right\rangle \left\langle {{s_4}} \right|} \right),{V_{1,2}} = \cos {\phi _ + }\left( {\left| {{s_1}} \right\rangle \left\langle {{s_3}} \right| + \left| {{s_4}} \right\rangle \left\langle {{s_2}} \right|} \right), \hfill \\
  {V_{2,1}} = \cos {\phi _ - }\left( {\left| {{s_3}} \right\rangle \left\langle {{s_2}} \right| + \left| {{s_1}} \right\rangle \left\langle {{s_4}} \right|} \right),{V_{2,2}} = \sin {\phi _ - }\left( {\left| {{s_1}} \right\rangle \left\langle {{s_3}} \right| - \left| {{s_4}} \right\rangle \left\langle {{s_2}} \right|} \right), \hfill \\ 
\end{gathered} \]
with 
\begin{equation}
{\phi _{+}}=\frac{{\theta +\varphi }}{2},{\phi _{-}}=\frac{{\theta -\varphi }%
}{2},
\end{equation}%
and the corresponding eigenfrequencies $\omega _{j\mu }$ as
\[{\omega _{11}} = {\omega _{21}} = \beta  - \alpha ,{\omega _{12}} = {\omega _{22}} = \beta  + \alpha. \]
Later we will use $\omega_\mu$ instead of $\omega_{j\mu}$ since $\omega_{j\mu}$ is independent of $j$ in our model. Based on the eigenoperators, one can rewrite 
 $\sigma _{j}^{x}=\sum\limits_{\mu }{({{V_{j\mu }}+V_{j\mu }^{\dag }})}$. Accordingly, the interaction Hamiltonian can also be rewritten as \begin{align}
{H_{S{B_{j}}}}\equiv B_{j}^{x}\sum\limits_{\mu }{\left( {{V_{j\mu }}+V_{j\mu }^{\dag }}\right).}
\end{align}
 
With the previous preliminary knowledge, one can directly get the master equation, subject to  the Born-Markov-Secular
approximation, as 
\begin{equation}
\frac{d{\rho }}{dt}=-i[{H}_{S}+H_{LS},{\rho }]+\mathcal{L}_{1}({\rho })+%
\mathcal{L}_{2}({\rho }),  \label{master}
\end{equation}%
where $\rho $ is the density matrix of the TLAs, $H_{LS}$ is the energy correction,
i.e., the Lamb shift, and $\mathcal{L}_{j}({\rho })$ are the dissipators
given by 
\begin{multline}
\mathcal{L}_{j}\left( \rho \right) =\sum\limits_{\mu =1}^{2}\left[{{\Gamma _{j}}%
\left( {{\omega _{\mu }}}\right) }\left( {{V_{j\mu }}\rho V_{j\mu }^{\dag
}-\frac{1}{2}\left\{ {\rho ,V_{j\mu }^{\dag }{V_{j\mu }}}\right\} }\right)
+{{\Gamma _{j}}\left( {-\omega _{\mu }}\right) }%
\left( {V_{j\mu }^{\dag }\rho {V_{j\mu }}-\frac{1}{2}\left\{ {\rho ,{%
V_{j\mu }}V_{j\mu }^{\dag }}\right\} }\right)\right] .  \label{Lj}
\end{multline}%
Here ${\Gamma _{j}}\left( \omega \right) =\int_{-\infty }^{\infty }{ds%
{e^{i\omega s}}\left\langle {B_{j}^{x}\left( s\right) B_{j}^{x}}%
\right\rangle }$ is the Fourier transform of the reservoir correlation
function $\left\langle {B_{j}^{x}\left( s\right) B_{j}^{x}}\right\rangle $,
so one can have 
\begin{gather}
\Gamma _{j}(\omega _{\mu })=2J_{j}(\omega _{\mu })[\bar{n}_{j}(\omega _{\mu
})+1], \\
\Gamma _{j}(-\omega _{\mu })=2J_{j}(\omega _{\mu })\bar{n}_{j}(\omega _{\mu
}),
\end{gather}%
where $\bar{n}_{j}(\omega )={(\exp(\beta _{j}\omega )-1)}^{-1}$ are the
average photon number and 
\begin{equation}
J_{j}\left( \omega \right) =\pi {\sum\limits_{n}{\left\vert {g_{j,n}}%
\right\vert }^{2}}\delta \left( {\omega -{\omega _{n}}}\right)
\end{equation}%
are the spectral densities of the heat reservoirs. In practical calculations, we typically replace the discrete sum over infinitely many delta-function-like modes with a continuous spectral density function. For instance, in this work, we consider an Ohmic-type thermal reservoir with a high-frequency cutoff $\omega _{D}$, whose spectral density takes the conventional form \cite{breuer2002theory,weiss2012quantum}: 
\begin{equation}
{J_{j}}\left( \omega \right) =\frac{{{\gamma _{j}}\omega }}{{1+{{\left( {%
\omega /{\omega _{D}}}\right) }^{2}}}}.  \label{density}
\end{equation}
This Drude cut-off provides a physically meaningful extension of the standard Ohmic model by incorporating a frequency-dependent damping term, thereby offering a more accurate representation of the underlying physical processes. This regularization scheme modifies the spectral density through a Lorentzian damping factor, which naturally introduces a smooth high-frequency cutoff. The resulting spectral density remains finite across all frequencies, resolving the unphysical divergence that occurs in the simple Ohmic case. This regularization is essential for constructing realistic models of quantum dissipation, as it properly accounts for the finite response times of physical environments while maintaining the characteristic linear frequency dependence at low energies. The cutoff frequency parameter simultaneously determines both the high-frequency roll-off and the characteristic timescale of environmental correlations.

The Lamb shift $H_{LS}$ in Eq. (\ref{master}) reads 
\begin{equation}
{H_{LS}}=\sum\limits_{j\mu }{\left[ {{S_{j}}\left( {\ {\omega _{\mu }}}%
\right) V_{j\mu }^{\dag }{V_{j\mu }}+{S_{j}}\left( {{-\omega _{\mu }}}%
\right) {V_{j\mu }}V_{j\mu }^{\dag }}\right] },
\end{equation}%
where 
\begin{gather}
{S_{j}}\left( {{\omega _{\mu }}}\right) =\frac{1}{\pi }P.V.\int\limits_{0}^{%
\infty }{{J_{j}}\left( \omega \right) \left( {\frac{{{{\bar{n}}_{j}}\left(
\omega \right) +1}}{{{\omega _{\mu }}-\omega }}+\frac{{{{\bar{n}}_{j}}\left(
\omega \right) }}{{{\omega _{\mu }}+\omega }}}\right) d\omega },\hfill
\label{S1} \\
{S_{j}}\left( {-{\omega _{\mu }}}\right) =-\frac{1}{\pi }P.V.\int%
\limits_{0}^{\infty }{{J_{j}}\left( \omega \right) \left( {\frac{{{{\bar{n}}%
_{j}}\left( \omega \right) }}{{{\omega _{\mu }}-\omega }}+\frac{{{{\bar{n}}%
_{j}}\left( \omega \right) +1}}{{{\omega _{\mu }}+\omega }}}\right) d\omega }%
,\hfill  \label{S2}
\end{gather}%
and $P.V.$ denotes the Cauchy principal value of the integral. The more
A detailed derivation is provided in Appendix \ref{appendixA}. Within our parameter range, we establish the following hierarchy of parameters:
\begin{equation}
    \gamma_j^{-1} \sim \omega_D/g \gg \omega_\mu/g \sim 1 \gg \gamma_j.
\end{equation}
Noting that the eigenfrequencies satisfies $\omega_2 - \omega_1 = 2\alpha \geq g \gg \gamma_j$. This establishes a crucial relationship between two characteristic timescales. The system's dynamical timescale $\tau_S = |\omega_1 - \omega_2|^{-1}$ is much shorter than the reservoir correlation time $\tau_R = \gamma_j^{-1}$, which ensures the validity of the secular approximation. Therefore, the global master equation approach is rigorously justified in this parameter regime, providing a consistent description of the open quantum system dynamics while properly accounting for the system's coherent evolution and dissipative processes.

\section{Lamb Shift}

\label{3}
Now, let's focus on the Lamb shift. Define 
\begin{gather}
{\Delta _{j\mu }}=\frac{{2{\omega _{\mu }}}}{\pi }P.V.\int_{0}^{\infty }{%
\frac{{{J_{j}}\left( \omega \right) {{\bar{n}}_{j}}\left( \omega \right) }}{{%
\omega _{\mu }^{2}-{\omega ^{2}}}}d\omega },\hfill  \label{D1} \\
\Delta _{j\mu }^{+}=\frac{1}{\pi }\int_{0}^{\infty }{\frac{{{J_{j}}\left(
\omega \right) }}{{{\omega _{\mu }}+\omega }}d\omega ,}\hfill \\
\Delta _{j\mu }^{-}=\frac{1}{\pi }P.V.\int_{0}^{\infty }{\frac{{{J_{j}}%
\left( \omega \right) }}{{{\omega _{\mu }}-\omega }}d\omega ,}\hfill
\end{gather}%
then the Lamb shift can be rewritten as 
\begin{equation}
{H_{LS}}=\sum\limits_{n=1}^{4}{{\Delta _{n}}\left\vert {s_{n}}\right\rangle
\left\langle {s_{n}}\right\vert }, \label{LS}
\end{equation}
where 
\begin{align}
    {\Delta _1} &=  - \left( {{\Delta _{1,1}} + \Delta _{1,1}^ + } \right){\sin ^2}{\phi _ + } - \left( {{\Delta _{2,2}} + \Delta _{2,2}^ + } \right){\sin ^2}{\phi _ - } \notag \\
   &- \left( {{\Delta _{1,2}} + \Delta _{1,2}^ + } \right){\cos ^2}{\phi _ + } - \left( {{\Delta _{2,1}} + \Delta _{2,1}^ + } \right){\cos ^2}{\phi _ - }, \hfill 
\end{align}
\begin{align}
  {\Delta _2} &= \left( {{\Delta _{1,1}} + \Delta _{1,1}^ - } \right){\sin ^2}{\phi _ + } + \left( {{\Delta _{2,2}} + \Delta _{2,2}^ - } \right){\sin ^2}{\phi _ - } \notag \\
   &+ \left( {{\Delta _{1,2}} + \Delta _{1,2}^ - } \right){\cos ^2}{\phi _ + } + \left( {{\Delta _{2,1}} + \Delta _{2,1}^ - } \right){\cos ^2}{\phi _ - }, \hfill 
\end{align}
\begin{align}
  {\Delta _3} &=  - \left( {{\Delta _{1,1}} + \Delta _{1,1}^ + } \right){\sin ^2}{\phi _ + } + \left( {{\Delta _{2,2}} + \Delta _{2,2}^ - } \right){\sin ^2}{\phi _ - } \notag \\
   &+ \left( {{\Delta _{1,2}} + \Delta _{1,2}^ - } \right){\cos ^2}{\phi _ + } - \left( {{\Delta _{2,1}} + \Delta _{2,1}^ + } \right){\cos ^2}{\phi _ - }, \hfill 
\end{align}
\begin{align}
  {\Delta _4} &= \left( {{\Delta _{1,1}} + \Delta _{1,1}^ - } \right){\sin ^2}{\phi _ + } - \left( {{\Delta _{2,2}} + \Delta _{2,2}^ + } \right){\sin ^2}{\phi _ - } \notag \\
   &- \left( {{\Delta _{1,2}} + \Delta _{1,2}^ + } \right){\cos ^2}{\phi _ + } + \left( {{\Delta _{2,1}} + \Delta _{2,1}^ - } \right){\cos ^2}{\phi _ - }.
\end{align}
It's easy to deduce that the Lamb shift commutes with the system's
Hamiltonian $H_{S}$. It acts as a modification to the energy levels of the
system. This implies that the Lamb shift doesn't alter the fundamental
structure of the Hamiltonian, but rather fine-tunes the energy levels within
the system, showcasing its role as a subtle yet pivotal adjustment factor in
open quantum system dynamics. From the equations above, we can obtain the
increments of the transition frequencies from $\left\vert s_{2}\right\rangle
\rightarrow \left\vert s_{3}\right\rangle $ and $\left\vert
s_{2}\right\rangle \rightarrow \left\vert s_{4}\right\rangle $ as 
\begin{eqnarray}
{\delta _{1}} =\Delta _{3}-\Delta _{2}=\left( {2{\Delta _{1,1}}+\Delta {_{1,1}^{\prime }}}\right) {\sin ^{2}}{%
\phi _{+}}+\left( {2{\Delta _{2,1}}+\Delta {_{2,1}^{\prime }}}\right) {\cos
^{2}}{\phi _{-}},  \label{25} \\
{\delta _{2}}=\Delta _{4}-\Delta _{2}=\left( {2{\Delta _{2,2}}+\Delta {_{2,2}^{\prime }}}\right) {\sin ^{2}}{%
\phi _{-}}+\left( {2{\Delta _{1,2}}+\Delta {_{1,2}^{\prime }}}\right) {\cos
^{2}}{\phi _{+}}, \label{26}
\end{eqnarray}%
where 
\begin{equation}
\Delta {_{j\mu }^{\prime }}=\Delta _{j\mu }^{+}+\Delta _{j\mu }^{-}=\frac{%
{2{\omega _{\mu }}}}{\pi }P.V.\int_{0}^{\infty }{\frac{{{J_{j}}\left( \omega
\right) }}{{\omega _{\mu }^{2}-{\omega ^{2}}}}d\omega }.  \label{D2}
\end{equation}

\begin{figure}[t]
\centering\includegraphics[width=9cm]{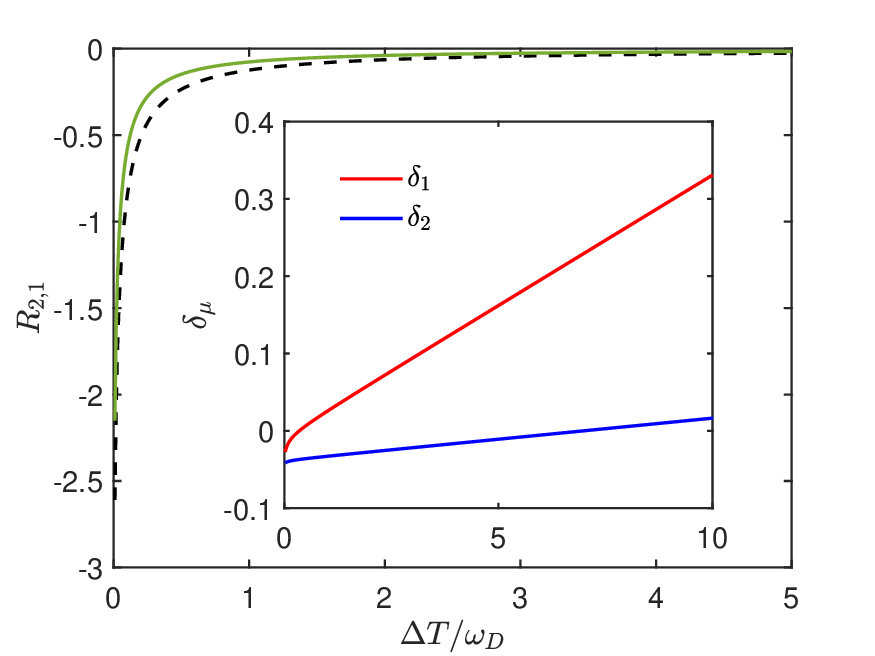}
\caption{${R_{2,1}}$ and its estimation vs. the relative temperature
difference $\Delta T/\protect\omega _{D}$. Here we set $T_{1}=1,T_{2}=1+%
\Delta T,\protect\gamma _{1}=\protect\gamma _{2}=0.01,\protect\omega _{D}=50,%
\protect\varepsilon _{1}=3,\protect\varepsilon _{2}=2,k=0.5.$ The green line
represents the estimation of ${R_{2,1}}$, the dashed black line represents
the exact value of ${R_{2,1}}$. For the inset, the red line represents $%
\protect\delta _{1}$ and the blue line represents $\protect\delta _{2}$ vs
the relative temperature difference $\Delta T/\protect\omega _{D}$ in the
same regime.}
\label{delta}
\end{figure}

To evaluate the effect of the Lamb shift, the critical task is to determine
the values of $\Delta _{j\mu }$ and $\Delta {_{j\mu }^{\prime }}$.
Fortunately, these values can be determined using the definition of the
Cauchy principal value integral 
\begin{equation}
P.V.\int\limits_{0}^{\infty }{\frac{{d\omega }}{{{\omega _{\mu }}-\omega }}=}%
\mathop {\lim }\limits_{\eta \rightarrow 0}\left( {\int_{0}^{{\omega _{\mu }}%
-\eta }{\frac{{d\omega }}{{{\omega _{\mu }}-\omega }}}+\int_{{\omega _{\mu }}%
+\eta }^{\infty }{\frac{{d\omega }}{{{\omega _{\mu }}-\omega }}}}\right) .
\end{equation}%
Applying the residue theorem, we have 
\begin{gather}
{\Delta _{j\mu }}=\frac{{{J_{j}}\left( {\omega _{\mu }}\right) }}{\pi }%
\left( {\ln \frac{{\omega _{D}}}{{\omega _{\mu }}}+\frac{\pi }{{{\beta _{j}}{%
\omega _{D}}}}+{R_{j\mu }}}\right) ,\hfill  \label{29} \\
\Delta {_{j\mu }^{\prime }}=-\frac{{2{J_{j}}\left( {\omega _{\mu }}\right) }%
}{\pi }\ln \frac{{\omega _{D}}}{{\omega _{\mu }}},\hfill  \label{30}
\end{gather}%
where 
\begin{equation}
{R_{j\mu }}=\frac{{2\pi }}{{\beta _{j}}}\sum\limits_{k=1}^{\infty }{\frac{{%
\omega _{\mu }^{2}-{\omega _{D}}{\omega _{k}}}}{{\left( {\omega _{\mu
}^{2}+\omega _{k}^{2}}\right) \left( {{\omega _{D}}+{\omega _{k}}}\right) }}}
\label{31}
\end{equation}
is a series with respect to $\omega _{k}$, and ${\omega _{k}}=2k\pi /{\beta
_{j}}$ is the Matsubara frequency. Using the Euler-Maclaurin formula, we can
obtain an estimation as 
\begin{equation}
{R_{j\mu }}=\ln \frac{\sqrt{4{\pi ^{2}}+\omega _{\mu }^{2}\beta _{j}^{2}}}{{%
2\pi +{\omega _{D}}{\beta _{j}}}}+O\left( {\beta _{j}}\right) .  \label{32}
\end{equation}%
The error of this estimation is clearly illustrated in Fig.\ref{delta}, and
we can observe a constant increasing $\delta _{\mu }$ as the temperature
increases from the inset. A more detailed discussion on this is provided in
Appendix \ref{appendixB}. Note that the negative initial value of $\delta _{\mu }$ suggests the Lamb shift's effect on heat current from reducing to enhancing contributions with increasing temperature difference.

\section{Heat Currents}

\label{4} We have obtained all the analytic expressions of the Lamb shift
and dissipators of the master equation \eqref{master}, hence we can get the
system's dynamics. Here, we are interested in the steady-state behavior of
the system. We have solved the density matrix of the system and found that
only the diagonal entries don't vanish so that we can express the steady-state
reduced density matrix as ${\rho}^{S}=diag[\rho_{11}^S,\rho_{22}^S,%
\rho_{33}^S,\rho_{44}^S],$ where \cite{sinaysky2008dynamics}
\begin{equation*}
\rho _{11}^S = \frac{{{X^ - }{Y^ - }}}{{XY}},\rho _{22}^S = \frac{{{X^ + }{%
Y^ + }}}{{XY}},\rho _{33}^S = \frac{{{X^ - }{Y^ + }}}{{XY}},\rho _{44}^S = 
\frac{{{X^ + }{Y^ - }}}{{XY}},
\end{equation*}
with 
\begin{align*}
&{X^ + } = {J_1}\left( {{\omega _1} }
\right){\bar{n}_1}\left( {{\omega _1} } \right){\sin ^2}{\phi _ + } +
{J_2}\left( {{\omega _1} } \right){\bar{n}_2}\left( {{\omega _1} }
\right){\cos ^2}{\phi _ - }, \\
&{Y^ + } = {J_1}\left( {{\omega _2} }
\right){\bar{n}_1}\left( {{\omega _2} } \right){\cos ^2}{\phi _ + } +
{J_2}\left( {{\omega _2} } \right){\bar{n}_2}\left( {{\omega _2} }
\right){\sin ^2}{\phi _ - }, \\
&{X^ - } = {J_1}\left( {{\omega _1} }
\right)\left( {{\bar{n}_1}\left( {{\omega _1} } \right) + 1} \right){\sin
^2}{\phi _ + } + {J_2}\left( {{\omega _1} } \right)\left(
{{\bar{n}_2}\left( {{\omega _1} } \right) + 1} \right){\cos ^2}{\phi _ - },\\ &{Y^ - } = {J_1}\left( {{\omega _2} } \right)\left( {{\bar{n}_1}\left(
{{\omega _2} } \right) + 1} \right){\cos ^2}{\phi _ + } +
{J_2}\left( {{\omega _2} } \right)\left( {{\bar{n}_2}\left( {{\omega _2} }
\right) + 1} \right){\sin ^2}{\phi _ - }, \\
&X = {X^ + } + {X^ - },Y = {Y^ +} + {Y^ - }.
\end{align*}
One can find that the above steady-state density matrix $\rho^S$ is the same
as the one without considering the Lamb shift. This is consistent with
the usual understanding that the Lamb shift does not affect the system's
eigenstates but only the eigenvalues.

To get the effect of the Lamb shift, we begin to study the heat current,
which is defined as \cite{breuer2002theory,e15062100} 
\begin{equation}
\mathcal{J}_{j}=Tr\left( {{({H}_{S}+{H}_{LS})}}\mathcal{L}_{j}({\rho }%
)\right) .  \label{J}
\end{equation}%
One can easily check that $[H_{S}+H_{LS},{\rho }^{S}]=0$, hence from the
master equation \eqref{master} we have 
\begin{equation}
\mathcal{L}_{1}({\rho }^{S})+\mathcal{L}_{2}({\rho }^{S})=\frac{d{\rho }^{S}%
}{dt}=0,
\end{equation}
which further implies that the two heat currents satisfy the conservation
relation $\mathcal{J}_{1}=-\mathcal{J}_{2}$. From Eq. (\ref{J}), we can give
the explicit form of the heat current as 
\begin{equation}
{\mathcal{J}_{1}^{\delta }}=\sum_{\mu=1}^{2}A_{\mu}\left( {{\bar{n}_{1}}\left( {
\omega _{\mu}}\right) -{\bar{n}_{2}}\left( {\omega _{\mu}}\right) }\right)
\left( {{\omega _{\mu}}+{\delta _{\mu}}}\right) \label{currentd}
\end{equation}
with 
\begin{gather}
A_{1}=2{\sin ^{2}}{\phi _{+}}{\cos ^{2}}{\phi _{-}}\frac{{{J_{1}}\left( {{%
\omega _{1}}}\right) {J_{2}}\left( {{\omega _{1}}}\right) }}{X},\hfill \\
A_{2}=2{\sin ^{2}}{\phi _{-}}{\cos ^{2}}{\phi _{+}}\frac{{{J_{1}}\left( {{%
\omega _{2}}}\right) {J_{2}}\left( {{\omega _{2}}}\right) }}{Y},\hfill
\end{gather}
and $\delta _{i}$ is from Eqs. \eqref{25} and \eqref{26}. We want to emphasize that $\delta $ in the heat current $%
\mathcal{J}_{1}^{\delta }$ is the signature of the Lamb shift, $\delta=\delta
_{i}=0 $ corresponding to ${\mathcal{J}_{1}^{0}}$ means that the Lamb shift
isn't considered. Thus, one can easily obtain the difference in heat
currents with and without considering the Lamb shift as 
\begin{equation}
{\Delta \mathcal{J}_{1}^{\delta }}=\sum_{\mu=1}^{2}A_{i}\left( {{n_{1}}\left( {\omega _{\mu}}\right) -{n_{2}}\left( {\omega _{\mu}}\right) }\right) {\delta _{\mu}}. \label{diff}
\end{equation}
Notice that $\mathcal{J}_{1}^{\delta }$ is always negative when $T_{2}>T_{1}$ from Eq. \eqref{56}, which is consistent with the second law of thermodynamics. There is no issue of the direction of heat current in this model because we're using the global approach for the master equation \cite{Levy_2014, Trushechkin_2016,Hofer_2017}. Next, we are only concerned with the magnitude of the heat currents, so the remainder of the discussion focuses on the absolute value of the heat currents.

\begin{figure}[t]
\centering\includegraphics[width=9cm]{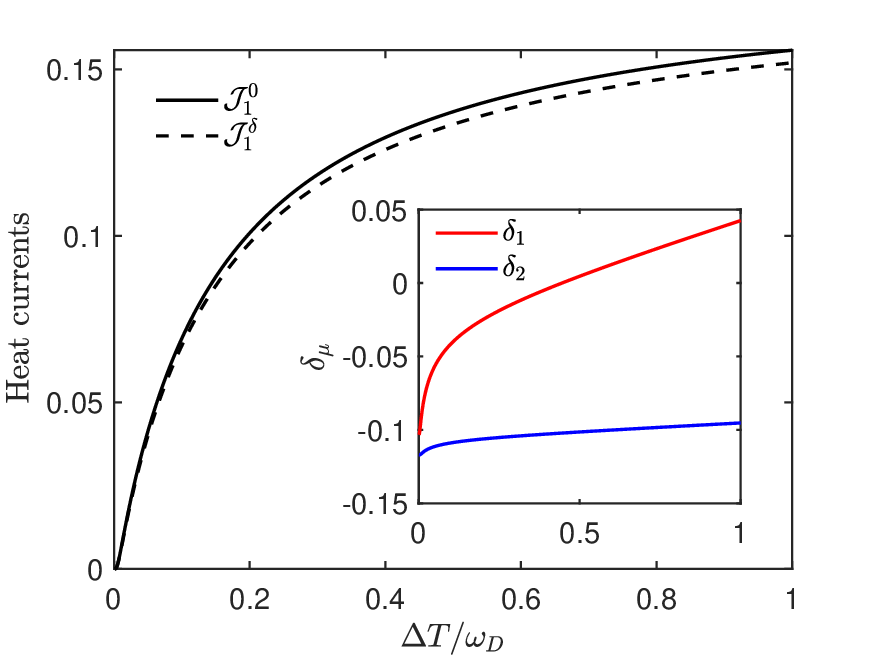}
\caption{The heat currents ${\mathcal{J}_{1}^{\protect\delta }}$ vs the
relative temperature difference $\Delta T/\protect\omega _{D}$. Here we set $\varepsilon _{1}=3,\varepsilon _{2}=2,T_{1}=0.1,T_{2}=0.1+\Delta T,\protect\gamma _{1}=\protect\gamma _{2}=0.02,\protect\omega _{D}=100,k=0.5.$ This figure illustrates the suppression of heat flow induced by the Lamb shift. For the inset, the red line represents $%
\protect\delta _{1}$ and the blue line represents $\protect\delta _{2}$ vs
the relative temperature difference $\Delta T/\protect\omega _{D}$ in the
same regime.}
\label{neg}
\end{figure}

From Eq. \eqref{diff}, it can be seen that when $\delta_{\mu}$ is negative, the Lamb shift has a suppressing effect on the heat current, as illustrated by Fig. \ref{neg}. Note that when $\Delta T>\omega_D/2$, $\delta_1$ is already positive, but since the reduction effect of the terms with $\delta_2$ is greater, $\Delta \mathcal{J}_{1}^{\delta }$ is still negative at this point. In fact, when the signs of $\delta_\mu$ are not the same, the size of $\Delta \mathcal{J}_{1}^{\delta }$ will depend on the competition between terms with $\delta_1$ and $\delta_2$.

From Eqs. \eqref{25}\eqref{26}\eqref{29}\eqref{30}, our analysis reveals that the value of omega exerts a substantial influence on delta. This can be attributed to the direct influence of the cutoff frequency on the Lamb shift, suggesting that the impact of the Lamb shift on the heat current is amplified with increasing cutoff frequency. Indeed, if the cutoff frequency tends to infinity, the Lamb shift will diverge.

\begin{figure}[t]
\centering\includegraphics[width=9cm]{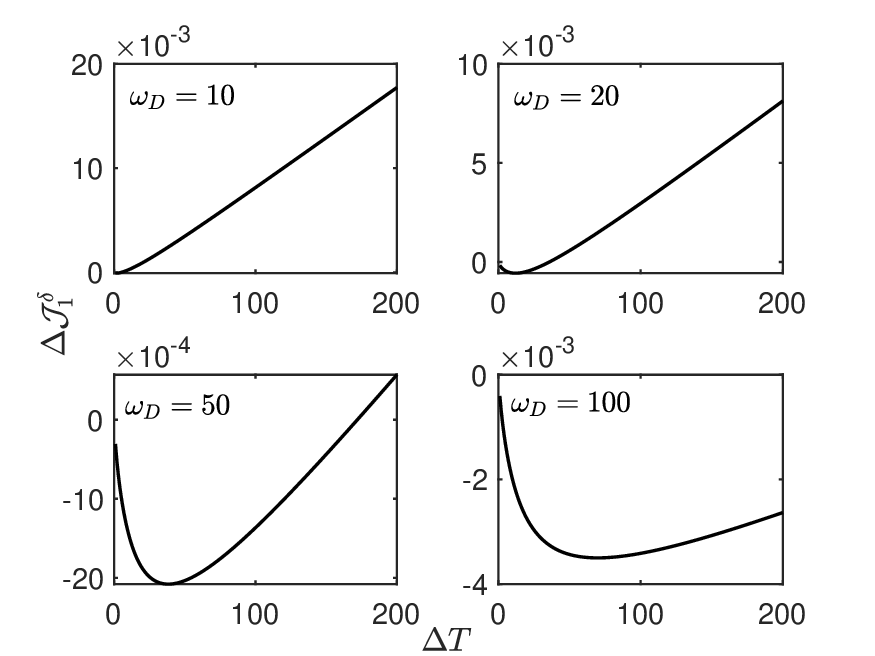}
\caption{The heat current difference $\Delta \mathcal{J}_{1}^{\delta }$ vs the temperature difference $\Delta T$. Here we set $\varepsilon _{1}=3,\varepsilon _{2}=2,T_{1}=1,T_{2}=1+\Delta T,\protect\gamma _{1}=\protect\gamma _{2}=0.02,k=0.5.$ For the insets, $\omega_D=10,20,50,100$, respectively. It can be seen that as $\omega_D$ increases, the temperature difference corresponding to the point where $\Delta \mathcal{J}_{1}^{\delta }$ changes from negative to positive becomes larger.}
\label{sub}
\end{figure}

Our analysis reveals that $\Delta \mathcal{J}_{1}^{\delta }$ exhibits a consistently developing linear dependence on large temperature difference, independent of the specific value of $\omega_D$, as illustrated by Fig. \ref{sub}. Namely, \textit{{when $T_{1}$ as the lower temperature is fixed the heat current ${\mathcal{J}_{1}^{0}}$ will approach an upper bound but ${\mathcal{J}_{1}^{\delta }}\rightarrow\infty $ with the temperature increment $\Delta T\rightarrow \infty $. }} To
show this, let's take the derivative of ${\mathcal{J}_{1}^{0}}$ with respect
to $\Delta T$, then we have 
\begin{align}
\frac{{d{\mathcal{J}_{1}^{0}}}}{{d\Delta T}} =2{K_{1}}\frac{{{J_{1}}\left( 
{\omega _{1}}\right) {{\sin }^{2}}{\phi _{+}}+{J_{2}}\left( {\omega _{1}}
\right) {{\cos }^{2}}{\phi _{-}}}}{{X^{2}}}{\omega _{1}} + 2{K_{2}}\frac{{{J_{2}}\left( {\omega _{2}}\right) {{\sin }^{2}}{\phi _{-}}
+{J_{1}}\left( {\omega _{2}}\right) {{\cos }^{2}}{\phi _{+}}}}{{Y^{2}}}{
\omega _{2}},
\end{align}
where 
\begin{align}
{K_{1}}& =\frac{{d{\bar{n}_{2}}\left( {\omega _{1}}\right) }}{{d\Delta T}}{%
\sin ^{2}}{\phi _{+}}{\cos ^{2}}{\phi _{-}}{J_{1}}\left( {\omega _{1}}%
\right) {J_{2}}\left( {\omega _{1}}\right) ({{2{\bar{n}_{1}}\left( {\omega
_{1}}\right) +1}}), \\
{K_{2}}& =\frac{{d{\bar{n}_{2}}\left( {\omega _{2}}\right) }}{{d\Delta T}}{%
\sin ^{2}}{\phi _{-}}{\cos ^{2}}{\phi _{+}}{J_{1}}\left( {\omega _{2}}%
\right) {J_{2}}\left( {\omega _{2}}\right) ({{2{\bar{n}_{1}}\left( {\omega
_{2}}\right) +1}}).
\end{align}%
Both $K_{1}$ and $K_{2}$ are two constant positive quantities. Thus, one can
easily obtain that ${\mathcal{J}_{1}^{0}}$ is a monotonically increasing
function of $\Delta T$. However, simple calculations can show that 
\begin{align}
\mathop {\lim }\limits_{\Delta T\rightarrow \infty }{A_{1}}{|\bar{n}_{1}({%
\omega _{1}})-\bar{n}_{2}({\omega _{1}})|}& ={J_{1}}\left( {\omega _{1}}%
\right) {\sin ^{2}}{\phi _{+}},  \label{42} \\
\mathop {\lim }\limits_{\Delta T\rightarrow \infty }{A_{2}}{|\bar{n}_{1}({%
\omega _{2}})-\bar{n}_{2}({\omega _{2}})|}& ={J_{1}}\left( {\omega _{2}}%
\right) {\cos ^{2}}{\phi _{+}},  \label{43}
\end{align}
and from Eq. (\ref{currentd}) we have 
\begin{equation}
\mathop {\lim }\limits_{\Delta T\rightarrow \infty }{\mathcal{J}_{1}^{0}}={%
J_{1}}\left( {\omega _{1}}\right) {\omega _{1}}{\sin ^{2}}{\phi _{+}}+{J_{1}}%
\left( {\omega _{2}}\right) {\omega _{2}}{\cos ^{2}}{\phi _{+}},
\label{upper}
\end{equation}
which remains constant since the system's parameters are fixed. Thus Eq. (%
\ref{upper}) serves as the supremum of ${\mathcal{J}_{1}^{0}}$. Namely,
without considering the Lamb shift, the heat current has an upper bound with the
temperature difference tending to infinity.

\begin{figure}[t]
\centering\includegraphics[width=9cm]{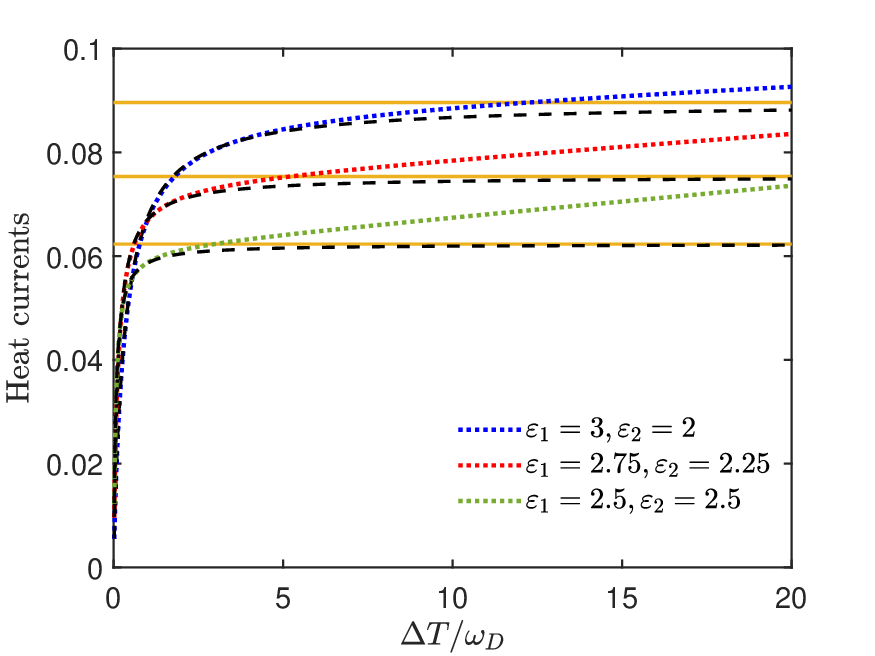}
\caption{The heat currents ${\mathcal{J}_{1}^{\protect\delta }}$ vs the
relative temperature difference $\Delta T/\protect\omega _{D}$. Here we set $%
T_{1}=1,T_{2}=1+\Delta T,\protect\gamma _{1}=\protect\gamma _{2}=0.01,%
\protect\omega _{D}=50,k=0.5.$ For the blue dotted line, $\protect%
\varepsilon _{1}=3,\protect\varepsilon _{2}=2;$ for the red dotted line, $%
\protect\varepsilon _{1}=2.75,\protect\varepsilon _{2}=2.25;$ for the green
dotted line, $\protect\varepsilon _{1}=2.5,\protect\varepsilon _{2}=2.5.$
The dashed black lines represent the heat currents $\mathcal{J}_{1}^{0}$ of
the same regime, but the Lamb shift is not taken into account, and the
orange solid lines represent the supremum of $\mathcal{J}_{1}^{0}$.}
\label{current}
\end{figure}

Now, let us turn to the case of the Lamb shift. Based on Eqs. (\ref{25})(\ref{26})(\ref{29})(\ref{30}), one find that $\delta _{\mu }$ can be rewritten
as 
\begin{equation}
{\delta _{\mu }}={P_{\mu }}+{Q_{\mu }}\Delta T+\frac{1}{\pi }{Q_{\mu }}{%
\omega _{D}}{R_{2,\mu }}, \label{estimation}
\end{equation}
where 
\begin{align}
& {P_{1}}=\frac{{2{J_{1}}\left( {{\omega _{1}}}\right) }}{\pi }\left( {\frac{%
\pi }{{{\beta _{1}}{\omega _{D}}}}+{R_{1,1}}}\right) {\sin ^{2}}{\phi _{+}}+%
\frac{{2{J_{2}}\left( {{\omega _{1}}}\right) }}{{{\beta _{1}}{\omega _{D}}}}{%
\cos ^{2}}{\phi _{-}},\hfill  \label{P1} \\
& {P_{2}}=\frac{{2{J_{1}}\left( {{\omega _{2}}}\right) }}{\pi }\left( {\frac{%
\pi }{{{\beta _{1}}{\omega _{D}}}}+{R_{1,2}}}\right) {\cos ^{2}}{\phi _{+}}+%
\frac{{2{J_{2}}\left( {{\omega _{2}}}\right) }}{{{\beta _{1}}{\omega _{D}}}}{%
\sin ^{2}}{\phi _{-}},\hfill  \label{P2}
\end{align}
and 
\begin{equation}
{Q_{1}}=\frac{{2{J_{2}}\left( {{\omega _{1}}}\right) }}{{{\omega _{D}}}}{%
\cos ^{2}}{\phi _{-}},{Q_{2}}=\frac{{2{J_{2}}\left( {{\omega _{2}}}\right) }%
}{{{\omega _{D}}}}{\sin ^{2}}{\phi _{-}}. \label{Q}
\end{equation}
Substituting Eqs. (\ref{42})(\ref{43})(\ref{estimation}) into Eq. (\ref{diff}), one can obtain 
\begin{align}
\mathop {\lim }\limits_{\Delta T\rightarrow \infty }\frac{{\Delta \mathcal{J}%
_{1}^{\delta }}}{{\Delta T}}={J_{1}}\left( {{\omega _{1}}}\right) {Q_{1}}%
\left( 1+\frac{{{\omega _{D}}}}{\pi }{\mathop {\lim }\limits_{\Delta
T\rightarrow \infty }\frac{{{R_{2,1}}}}{{\Delta T}}}\right){\sin ^{2}}{\phi
_{+}}\hfill  \notag \\
+{J_{1}}\left( {{\omega _{2}}}\right) {Q_{2}}\left( {1+\frac{{{\omega _{D}}%
}}{\pi }\mathop {\lim }\limits_{\Delta T\rightarrow \infty }\frac{{{R_{2,2}}}%
}{{\Delta T}}}\right) {\cos ^{2}}{\phi _{+}}. \label{49}
\end{align}
Considering Eqs. (\ref{31}) and (\ref{32}), we have 
\begin{equation}
\mathop {\lim }\limits_{\Delta T\rightarrow \infty }{R_{2,\mu }}/\Delta T=0.
\label{part}
\end{equation}
Substituting Eq. (\ref{part}) into Eq. (\ref{49}), we will arrive at 
\begin{equation}
\mathop {\lim }\limits_{\Delta T\rightarrow \infty }\frac{{\Delta \mathcal{J}%
_{1}^{\delta }}}{{\Delta T}}={J_{1}}\left( {{\omega _{1}}}\right) {Q_{1}}{%
\sin ^{2}}{\phi _{+}}+{J_{1}}\left( {{\omega _{2}}}\right) {Q_{2}}{\cos ^{2}}%
{\phi _{+}}.
\end{equation}
This indicates that the $\Delta \mathcal{J}_{1}^{\delta }$ increases
linearly with the temperature difference in the regime $\Delta T\rightarrow
\infty $. That is, the heat current with Lamb shift in the regime $\Delta
T\rightarrow \infty $, the sum of the $\mathcal{J}_{1}^0$ and $\Delta\mathcal{J}_{1}^{\delta }$, can exceed the upper bound Eq. (\ref{upper}) due to the linearly increasing $\Delta \mathcal{J}_{1}^{\delta }$. Such a result is also explicitly illustrated in Fig. \ref{current}. It can be seen that at slight temperature differences, the dashed line is below the dotted line, but the difference is not significant. With the temperature increasing, all the dashed lines are
tightly below the corresponding orange solid line. In contrast, the dotted lines
increase linearly and exceed the supremum corresponding to the orange solid
line. In addition, it can be observed that when $\varepsilon
_{1}+\varepsilon _{2}$ is a constant, the smaller the $\left\vert
\varepsilon _{1}-\varepsilon _{2}\right\vert $ is, the smaller the temperature difference is at which the heat current with Lamb shift surpasses the
supremum of heat current corresponding to the orange solid line.

We have chosen $J^{(1)}(\omega )=\frac{\gamma \omega }{1+(\omega /\omega
_{D})^{2}}$ as the spectral density of the reservoirs in the previous study.
We can also choose a simple, discontinuous cut-off as 
\begin{equation}
\begin{aligned} &J^{(2)}(\omega)=\gamma
\omega,\omega<\omega_D,\\&J^{(2)}(\omega)=0,\omega\geq\omega_D; \end{aligned}
\end{equation}
or a cut-off for exponential decay as 
\begin{equation}
J^{(3)}(\omega )=\gamma \omega exp(-\omega ^{2}/\omega _{D}^{2}).
\end{equation}
Notice that all three represent the Ohmic-type heat reservoir.
Furthermore, since $\omega _{\mu }\ll \omega _{D},$ $J^{(1)}(\omega _{\mu
})-J^{(2)}(\omega _{\mu })$ and $J^{(3)}(\omega _{\mu })-J^{(2)}(\omega
_{\mu })$ are equivalent infinitesimals of the same order $O(\omega _{\mu
}^{2}/\omega _{D}^{2})$. Therefore, the results obtained without the Lamb
Shift should be very similar, as indicated by the dashed lines in Fig. \ref%
{j1j2j3}. We can also see that, considering the Lamb Shift, the influence of $%
J^{(1)}$ on the heat current lies between $J^{(2)}$ and $J^{(3)}$, and the
different Lamb Shifts corresponding to these three spectral densities, all
lead to a linear increase in $\delta _{\mu }$ with the increasing
temperature difference, eventually resulting in a linear increase in the
heat current. 
\begin{figure}[t]
\centering\includegraphics[width=9cm]{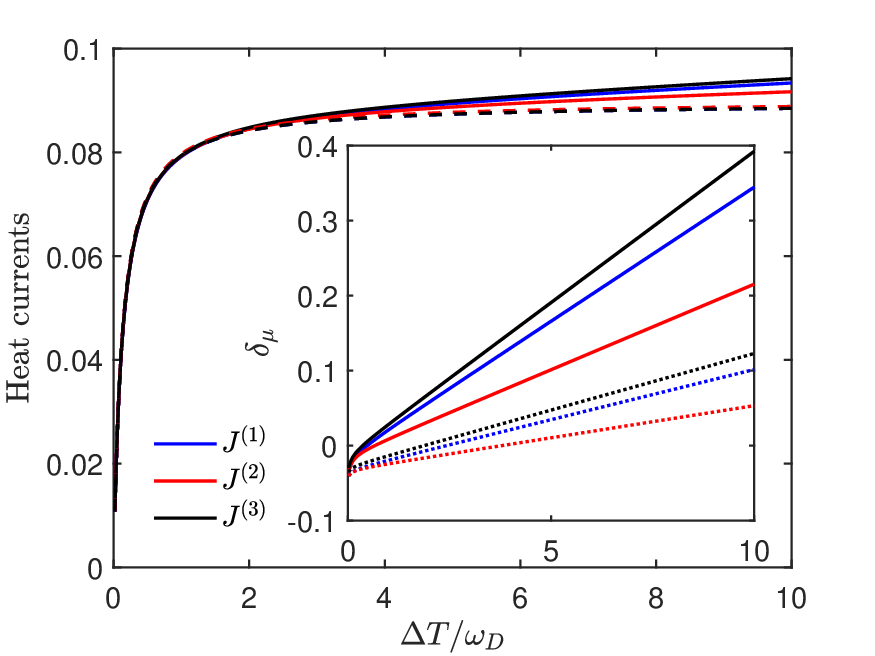}
\caption{The heat currents ${\mathcal{J}_{1}^{\protect\delta }}$ vs the
relative temperature difference $\Delta T/\protect\omega _{D}$. Here we set $%
T_{1}=1,T_{2}=1+\Delta T,\protect\gamma _{1}=\protect\gamma _{2}=0.01,%
\protect\omega _{D}=50,\protect\varepsilon _{1}=3,\protect\varepsilon %
_{2}=2.5,k=0.5.$ For the blue line, $J^{(1)}(\protect\omega )=\frac{\protect%
\gamma \protect\omega }{1+(\protect\omega /\protect\omega _{D})^{2}};$ for
the red line, $J^{(2)}(\protect\omega )=\protect\gamma \protect\omega $ when 
$\protect\omega <\protect\omega _{D}$, $J^{(2)}(\protect\omega )=0$ when $%
\protect\omega \geq \protect\omega _{D};$ for the black line, $J^{(3)}(%
\protect\omega )=\protect\gamma \protect\omega exp(-\protect\omega ^{2}/%
\protect\omega _{D}^{2}).$ The dashed lines represent the heat currents $%
\mathcal{J}_{1}^{0}$ of the same regime, but the Lamb shift is not
considered. For the inset, the solid lines represent $\protect\delta _{1}$,
and the dotted lines represent $\protect\delta _{2}$ vs the relative
temperature difference $\Delta T/\protect\omega _{D}$ in the same regime;
The correspondence of colors remains consistent with the previous text.}
\label{j1j2j3}
\end{figure}

\section{Conclusions}

\label{6}
We investigate the influence of the Lamb shift on heat transport in a two-qubit system coupled to thermal reservoirs at different temperatures. Our results demonstrate a dual role of the Lamb shift in regulating heat current: 
\begin{itemize}
\item At small temperature differences, the Lamb shift suppresses the steady-state heat current; 
\item In contrast, for large temperature differences, the system exhibits distinct behavior - while the heat current saturates to an upper bound when neglecting the Lamb shift, its inclusion leads to a divergent heat current as the temperature gradient approaches infinity.
\end{itemize}

Our findings yield profound insights into the fundamental mechanisms of quantum heat transport, revealing how quantum coherence and system-reservoir interactions collectively govern heat current at the quantum level. It should be noted that the observed phenomenon tends to be less pronounced under small temperature differences. Future work could extend this investigation to the case with stronger temperature gradients, where the applicability of the master equation and the possible influence of non-Markovian effects may offer interesting avenues. The revealed modification of heat current induced by Lamb shift suggests new strategies for controlling heat current through quantum engineering of system-environment interactions. This work lays the foundation for future studies of quantum heat engines, quantum batteries, and the development of quantum materials with tailored thermal properties. Furthermore, the broader interest of the Lamb shift could also be cast to quantum thermodynamics and open quantum systems, such as the quantum description of Otto cycles \cite{thomas2011coupled,kosloff2017quantum,PhysRevA.107.032220,e22111255,kloc2019collective} and q-deformation in heat engines \cite{ozaydin2023powering,guvenilir2022work,xiao2023performance}. In addition, one could  explore the implications of the Lamb shift in cavity quantum electrodynamics (QED) \cite{PhysRevLett.118.213601,ladd2006hybrid,agarwal2015photon,PhysRevA.90.053811}, particularly in micromaser systems. When a cavity mode repeatedly interacts with a stream of atomic ensembles, the Lamb shift could significantly influence the system's dynamics, the onset of maser action, and the resulting photon statistics. A detailed investigation into this effect could provide deeper insight into the role of vacuum fluctuations and field-atom interactions in non-equilibrium quantum systems, thereby bridging our findings with ongoing experimental efforts in cavity-based quantum information processing.





\funding{This work was supported by the National Natural Science Foundation of China under Grants Nos. 12575009 and 12175029.}

\dataavailability{The original contributions presented in this study are included in the article. Further inquiries can be directed to the corresponding author.}

\conflictsofinterest{The authors declare no conflicts of interest.}



\appendixtitles{yes} 
\appendixstart
\appendix

\section{The Derivation of the Master Equation}

\label{appendixA}
In the interaction picture, the dynamic evolution of the
the von Neumann equation describes the whole system plus the environment 
\begin{equation}
\frac{{d\rho }^{\prime }{\left( t\right) }}{{dt}}=-i\left[ {{H_{I}}\left(
t\right) ,\rho }^{\prime }{\left( t\right) }\right] .
\end{equation}%
where ${\rho }^{\prime }{\left( t\right) }$ denotes the total density
matrix. With the standard procedure of Born-Markov-Secular approximation, we
can get the evolution of the reduced density matrix ${\rho \left( t\right) }$
as 
\begin{align}
\frac{{d\rho \left( t\right) }}{{dt}}=\sum\limits_{\mu ,i,j}{{C_{i,j}}%
\left( {\omega _{\mu }}\right) \left[ {{V_{j\mu }}\rho \left( t\right)
,V_{i,\mu }^{\dag }}\right] }+\sum\limits_{\mu ,i,j}{C_{j,i}^{\ast }\left( {\omega _{\mu }}\right) %
\left[ {{V_{j\mu }},\rho \left( t\right) V_{i,\mu }^{\dag }}\right] },
\end{align}
where 
\begin{equation}
{C_{i,j}}\left( {\omega _{\mu }}\right) =\int\limits_{0}^{\infty }{ds{e^{i{%
\omega _{\mu }}s}}\left\langle {B_{i}^{x}\left( s\right) B_{j}^{x}}%
\right\rangle }
\end{equation}%
is the forward Fourier transform of the reservoir correlation function $%
\left\langle {B_{i}^{x}\left( s\right) B_{j}^{x}}\right\rangle $. Actually
only the terms $\left\langle {B_{j}^{x}\left( s\right) B_{j}^{x}}%
\right\rangle $ are not zero, because 
\begin{align}
Tr\left( {B_{j}^{x}\left( s\right) B_{j}^{x}{\rho _{B}}}\right)=\sum\limits_{n}{{{\left\vert {{g_{j,n}}}\right\vert }^{2}}\left( {{e^{-i{%
\omega _{n}}s}}\left( {{{\bar{n}}_{j}}\left( {\omega _{n}}\right) +1}\right)
+{e^{i{\omega _{n}}s}}{{\bar{n}}_{j}}\left( {\omega _{n}}\right) }\right) }%
\hfill  \notag \\
=\frac{1}{\pi }\int\limits_{0}^{\infty }{{J_{j}}\left( \omega \right)
\left( {{e^{-i{\omega }s}}\left( {{{\bar{n}}_{j}}\left( {\omega }\right) +1}%
\right) +{e^{i{\omega }s}}{{\bar{n}}_{j}}\left( {\omega }\right) }\right)
d\omega },\hfill
\end{align}%
so we can obtain that 
\begin{align}
& {C_{j,j}}\left( {\omega _{\mu }}\right) =\int\limits_{0}^{\infty }{ds{e^{i{%
\omega _{\mu }}s}}Tr\left( {B_{j}^{x}\left( s\right) B_{j}^{x}{\rho _{B}}}%
\right) }\hfill  \notag \\
& =\frac{1}{\pi }\int\limits_{0}^{\infty }{\left( {\int\limits_{0}^{\infty }{%
{e^{-i\left( {\omega -{\omega _{\mu }}}\right) s}}ds}}\right) {J_{j}}\left(
\omega \right) \left( {{{\bar{n}}_{j}}\left( \omega \right) +1}\right) }%
d\omega  \notag \\
& +\frac{1}{\pi }\int\limits_{0}^{\infty }{\left( {\int\limits_{0}^{\infty }{%
{e^{-i\left( {\ -{\omega _{\mu }}-\omega }\right) s}}ds}}\right) {J_{j}}%
\left( \omega \right) {{\bar{n}}_{j}}\left( \omega \right) }d\omega .
\end{align}%
Using the Kramers-Kronig relations 
\begin{equation}
\int\limits_{0}^{\infty }{{e^{-i\left( {\omega -{\omega _{0}}}\right) s}}ds}%
=\pi \delta \left( {\omega -{\omega _{0}}}\right) -iP.V.\frac{{1}}{{\omega -{%
\omega _{0}}}},
\end{equation}%
and by the properties of the delta function
\begin{align}
\int\limits_0^\infty  {\delta \left( {\omega  - {\omega _\mu }} \right){J_j}\left( \omega  \right)\left( {{{\overline n }_j}\left( \omega  \right) + 1} \right)d\omega }  = {J_j}\left( {{\omega _\mu }} \right)\left( {{{\overline n }_j}\left( {{\omega _\mu }} \right) + 1} \right),
\end{align}
\begin{align}
\int\limits_0^\infty  {\delta \left( {\omega  + {\omega _\mu }} \right){J_j}\left( \omega  \right){{\overline n }_j}\left( \omega  \right)d\omega }  = 0,
\end{align}
we finally arrived at 
\begin{align}
{C_{j,j}}\left( {\omega _{\mu }}\right) ={J_{j}}\left( {\omega _{\mu }}%
\right) \left( {{\bar{n}}_{j}\left( {\omega _{\mu }}\right) +1}\right)+\frac{i}{\pi }P.V.\int_{0}^{\infty }{{J_{j}}\left( \omega \right) \left( {%
\frac{{{{\bar{n}}_{j}}\left( \omega \right) +1}}{{{\omega _{\mu }}-\omega }}+%
\frac{{{{\bar{n}}_{j}}\left( \omega \right) }}{{{\omega _{\mu }}+\omega }}}%
\right) d\omega,}\hfill
\end{align}
where the $P.V.$ denotes the Cauchy principal value. So we've got the real
parts $\frac{1}{2}{\Gamma _{j}}\left( {\omega _{\mu }}\right) $ and
imaginary parts ${S_{j}}\left( {\omega _{\mu }}\right) $ of ${C_{j,j}}({%
\omega _{\mu }})$. Based on the same method, we can obtain the real parts $%
\frac{1}{2}{\Gamma _{j}}\left( {-\omega _{\mu }}\right) $ and imaginary
parts ${S_{j}}\left( {-\omega _{\mu }}\right) $ of ${C_{j,j}}({-\omega _{\mu
}})$ as 
\begin{align}
{C_{j,j}}\left( {\ -{\omega _{\mu }}}\right) ={J_{j}}\left( {\omega _{\mu }%
}\right) \bar{n}\left( {\omega _{\mu }}\right) -\frac{i}{\pi }P.V.\int_{0}^{\infty }{{J_{j}}\left( \omega \right) \left( {%
\frac{{{{\bar{n}}_{j}}\left( \omega \right) }}{{{\omega _{\mu }}-\omega }}+%
\frac{{{{\bar{n}}_{j}}\left( \omega \right) +1}}{{{\omega _{\mu }}+\omega }}}%
\right) d\omega .}
\end{align}%
Note that the frequency domain KMS condition is satisfied, i.e., 
\begin{equation}
{\Gamma _{j}}\left( {\ -\omega }\right) ={e^{-\beta \omega }}{\Gamma _{j}}%
\left( \omega \right) .
\end{equation}

\section{Details of the Estimate of the Lamb Shift}

\label{appendixB}
The two most critical integrals for obtaining the Lamb Shift (\ref{LS}) in this paper are Eqs.(\ref{D1})(\ref{D2}), if we put 
\begin{equation}
{f_{j\mu }}\left( \omega \right) = \frac{\omega}{{\omega _D^2 + {\omega ^2}}%
}\frac{{{{\bar{n} }_j}\left( \omega \right)}}{{{\omega _\mu } - \omega }},
\end{equation}
\begin{equation}
{F_{j\mu }}\left( \omega \right) = \frac{\omega}{{\omega _D^2 + {\omega ^2}}%
}\frac{{{{\bar{n} }_j}\left( \omega \right)}}{{{\omega _\mu } + \omega }},
\end{equation}
then $f_{j\mu }$ has only one positive pole on the real axis, and $F_{j\mu
}$ has only one negative pole on the real axis. So the problem is to find the principal integrals of these two functions on the positive real axis, using the residue theorem (and lots of contours), we find that 
\begin{equation}
\begin{gathered} P.V.\int\limits_0^\infty {{f_{j\mu }}\left( \omega
\right)d\omega } = \operatorname{Res} \left[ {f_{j\mu }\left( { - \omega }
\right)\ln \left( \omega \right),{\omega _D}i} \right] +
\operatorname{Res} \left[ {f_{j\mu }\left( { - \omega } \right)\ln \left( \omega
\right), - {\omega _D}i} \right] \\  + \sum\limits_{k = 1}^\infty
{\operatorname{Res} \left[ {f_{j\mu }\left( { - \omega } \right)\ln \left( \omega
\right),{\omega _k}i} \right]}  + \sum\limits_{k = 1}^\infty
{\operatorname{Res} \left[ {f_{j\mu }\left( { - \omega } \right)\ln \left( \omega
\right), - {\omega _k}i} \right]} - \operatorname{Res} \left[
{f_{j\mu }\left( \omega \right)\ln \left( \omega \right),{\omega _\mu }} \right],
 \end{gathered}
\end{equation}
and
\begin{equation}
\begin{gathered} - 2\pi \int\limits_0^\infty {{F_{j\mu }}\left( \omega
\right)d\omega } = \operatorname{Im} \left( {\operatorname{Res}
\left[ {F_{j\mu }\left( \omega \right){{\ln }^2}\left( \omega \right),{\omega _D}i}
\right]} \right) + \operatorname{Im} \left( {\operatorname{Res}
\left[ {F_{j\mu }\left( \omega \right){{\ln }^2}\left( \omega \right), - {\omega
_D}i} \right]} \right) \\ + \operatorname{Im} \sum\limits_{k =
1}^\infty {\operatorname{Res} \left[ {F_{j\mu }\left( \omega \right){{\ln }^2}\left(
\omega \right),{\omega _k}i} \right]}  +\operatorname{Im}
\sum\limits_{k = 1}^\infty {\operatorname{Res} \left[ {F_{j\mu }\left( \omega
\right){{\ln }^2}\left( \omega \right), - {\omega _k}i} \right]} \\+
\operatorname{Im} \left( {\operatorname{Res} \left[ {F_{j\mu }\left( \omega
\right){{\ln }^2}\left( \omega \right), - {\omega _\mu }} \right]} \right), \end{gathered}
\end{equation}
that is
\begin{align}
P.V.\int\limits_0^\infty {{f_{j\mu }}\left( \omega \right)d\omega } = \frac{{{\omega _\mu }\ln {\omega _\mu }}}{{\omega _D^2 + \omega _\mu ^2}}%
\frac{1}{{\exp \left( {{\beta _j}{\omega _\mu }} \right) - 1}} + \frac{1}{2}%
\frac{{{\omega _\mu }\ln {\omega _D} + \pi {\omega _D}/2}}{{\omega _D^2 +
\omega _\mu ^2}}  \notag \\
+ \frac{1}{{\beta _j}}\sum\limits_{k = 1}^\infty {\frac{{{\omega _k}\left( {%
2{\omega _k}\ln {\omega _k} - \pi {\omega _\mu }} \right)}}{{\left( {\omega
_k^2 + \omega _\mu ^2} \right)\left( {\omega _D^2 - \omega _k^2} \right)}}}- \frac{1}{2}\frac{{{\omega _D}\ln {\omega _D} - \pi {\omega _\mu }/2}}{{%
\omega _D^2 + \omega _\mu ^2}}\cot \left( {{\beta _j}{\omega _D}/2} \right),
  \label{f}
\end{align}
and
\begin{align}
\int\limits_0^\infty {{F_{j\mu }}\left( \omega \right)d\omega }  
= \frac{{{\omega _\mu }\ln {\omega _\mu }}}{{\omega _D^2 + \omega _\mu ^2}}%
\frac{1}{{\exp \left( {\ - {\beta _j}{\omega _\mu }} \right) - 1}} + \frac{1%
}{2}\frac{{{\omega _\mu }\ln {\omega _D} - \pi {\omega _D}/2}}{{\omega _D^2
+ \omega _\mu ^2}} \hfill  \notag \\
- \frac{1}{{\beta _j}}\sum\limits_{k = 1}^\infty {\frac{{{\omega _k}\left( {%
2{\omega _k}\ln {\omega _k} + \pi {\omega _\mu }} \right)}}{{\left( {\omega
_k^2 + \omega _\mu ^2} \right)\left( {\omega _D^2 - \omega _k^2} \right)}}} 
+ \frac{1}{2}\frac{{{\omega _D}\ln {\omega _D} + \pi {\omega _\mu }/2}}{{%
\omega _D^2 + \omega _\mu ^2}}\cot \left( {{\beta _j}{\omega _D}/2} \right). \label{F}
\end{align}
Here ${\omega _k} = \frac{{2k\pi }}{{{\beta _j}}} = k{\omega _1}$ is the module of the poles on the imaginary axis.

Thus we have
\begin{align}
&\frac{\pi }{{{\gamma _j}\omega _D^2}}{\Delta _{j\mu }} =
\int\limits_0^\infty {\left( {{f_{j\mu }}\left( \omega \right) + {F_{j\mu }%
}\left( \omega \right)} \right)d\omega } \notag \\
&= \frac{{\omega _\mu }}{{\omega _D^2 + \omega _\mu ^2}}\left( {\ln \frac{{%
\omega _D}}{{\omega _j}} + \frac{\pi }{2}\cot \frac{{{\beta _j}{\omega _D}}}{%
2}} \right)
- \frac{{2\pi {\omega _\mu }}}{{\beta _j}}\sum\limits_{k = 1}^\infty {\frac{%
{\omega _k}}{{\left( {\omega _k^2 + \omega _\mu ^2} \right)\left( {\omega
_D^2 - \omega _k^2} \right)}}} . \label{De}
\end{align}
$\Delta _{j\mu}$ contains the cotangent
function, where an apparent divergence emerges as $\beta _j$ approaches certain critical values. Contrary to initial indications, however, such a divergence
is eliminated due to a precise cancellation of divergent terms within the
series in $\Delta _{j\mu}$. This behavior is attributed to the
convergence of a movable pole $\omega_ki$ toward a stationary pole $%
\omega_Di $. At these specific values of $\beta _j$, the coincidence of the
two poles result in the formation of a second-order pole. In other words,
as a function of both $\beta _j$ and $\omega_D$, $\Delta _{j\mu}$ remains
continuous. This continuity also aligns with physical intuition: the Lamb
shift, as a correction to the system's energy levels, does not exhibit
divergence as the temperature and the cut-off frequency undergo continuous
variations. This fact can also be clarified using the Mittag-Leffler
expansion of the cotangent function 
\begin{equation}
\cot \left( x \right) = \frac{1}{x} - 2x\sum\limits_{k = 1}^\infty {\frac{1}{%
{{{\left( {k\pi } \right)}^2} - {x^2}}}},  \label{cot}
\end{equation}
substituting this into \eqref{De}, we arrive at 
\begin{equation}
\frac{{\pi {\Delta _{j\mu }}}}{{{J_j}\left( {\omega _\mu } \right)}} = \ln 
\frac{{\omega _D}}{{\omega _\mu }} + \frac{\pi }{{{\beta _j}{\omega _D}}} + 
\frac{{2\pi }}{{\beta _j}}\sum\limits_{k = 1}^\infty {\frac{{\omega _\mu ^2
- {\omega _D}{\omega _k}}}{{\left( {\omega _\mu ^2 + \omega _k^2}
\right)\left( {{\omega _D} + {\omega _k}} \right)}}}. \label{B7}
\end{equation}

More directly if we only care about $\delta_\mu$ in Eqs.(\ref{25})(\ref{26}), we only need to calculate 
\begin{equation}
2{\Delta _{j\mu }} + \Delta {'_{j\mu }} = \frac{{2{\omega _\mu }}}{\pi }P.V.\int\limits_0^\infty  {\frac{{{J_j}\left( \omega  \right)}}{{\omega _\mu ^2 - {\omega ^2}}}\coth \left( {\frac{{\beta_j \omega }}{2}} \right)d\omega },
\end{equation}
Using the Mittag-Leffler expansion of the hyperbolic cotangent function 
\begin{equation}
\coth \left( {\frac{{\beta_j \omega }}{2}} \right) = \frac{2}{{\beta_j \omega }} + \frac{2}{\beta_j }\sum\limits_{k = 1}^\infty  {\frac{{2\omega }}{{{\omega ^2} + \omega _k^2}}},
\end{equation}
with the residue theorem, we have 
\begin{align}
&\frac{{4{\omega _\mu }}}{{\beta_j \pi }}P.V.\int\limits_0^\infty  {\frac{{{J_j}\left( \omega  \right)}}{{\omega _\mu ^2 - {\omega ^2}}}\frac{1}{\omega }d\omega }  = {J_j}\left( {{\omega _\mu }} \right)\frac{2}{{\beta_j {\omega _D}}} \hfill \\
&\frac{{8{\omega _\mu }}}{{\beta_j \pi }}P.V.\int\limits_0^\infty  {\frac{{{J_j}\left( \omega  \right)}}{{\omega _\mu ^2 - {\omega ^2}}}\frac{\omega }{{{\omega ^2} + \omega _k^2}}d\omega } = {J_j}\left( {{\omega _\mu }} \right)\frac{4}{\beta_j }\frac{{\omega _\mu ^2 - {\omega _D}{\omega _k}}}{{\left( {{\omega _D} + {\omega _k}} \right)\left( {\omega _\mu ^2 + \omega _k^2} \right)}} \hfill
\end{align}
Thus 
\begin{equation}
\frac{{2{\Delta _{j\mu }} + \Delta {'_{j\mu }}}}{{{J_j}\left( {{\omega _\mu }} \right)}} = \frac{2}{{\beta_j {\omega _D}}} + \frac{4}{\beta_j }\sum\limits_{k = 1}^\infty  {\frac{{\omega _\mu ^2 - {\omega _D}{\omega _k}}}{{\left( {{\omega _D} + {\omega _k}} \right)\left( {\omega _\mu ^2 + \omega _k^2} \right)}}},
\end{equation}
which is consistent with Eq.(\ref{B7}).

In order to estimate the residual term $R_{j\mu }$, we set 
\begin{align}
G\left( k \right) &= \frac{{\omega _\mu ^2 - {\omega _D}{\omega _k}}}{{%
\left( {\omega _\mu ^2 + \omega _k^2} \right)\left( {{\omega _D} + {\omega _k%
}} \right)}} = \frac{{\omega _\mu ^2 - k{\omega _1}{\omega _D}}}{{\left( {\omega _\mu ^2
+ {k^2}\omega _1^2} \right)\left( {{\omega _D} + k{\omega _1}} \right)}}.
\end{align}
Notice that the ${\omega _1} = {\left. {{\omega _k}} \right|_{k = 1}}$ here is not the same as the ${\omega _1} = {\left. {{\omega _\mu }} \right|_{\mu  = 1}}$ in the main text, using the Euler-Maclaurin formula 
\begin{equation}
\begin{gathered} \sum\limits_{k = 1}^n {G\left( k \right)} - \int_0^n
{G\left( x \right)dx} - \frac{{G\left( n \right)}}{2} - \sum\limits_{k =
1}^\infty {\frac{{{B_{2k}}}}{{\left( {2k} \right)!}}{G^{\left( {2k - 1}
\right)}}\left( n \right)} \hfill \\ = \int_1^0 {G\left( x \right)dx} +
\frac{{G\left( 1 \right)}}{2} - \sum\limits_{k = 1}^\infty
{\frac{{{B_{2k}}}}{{\left( {2k} \right)!}}{G^{\left( {2k - 1}
\right)}}\left( 1 \right)} , \hfill \end{gathered}
\end{equation}
where $B_{2k}$ is the $2k$th Bernoulli number, letting $n$ approach infinity,
we get 
\begin{align}
{\omega _1}\sum\limits_{k = 1}^n {G\left( k \right)} - \ln \frac{{\omega
_\mu }}{{\omega _D}} &= \frac{1}{2}\ln \frac{{\omega _D^2\left( {\omega _1^2 + \omega _\mu ^2}
\right)}}{{\omega _\mu ^2{{\left( {{\omega _1} + {\omega _D}} \right)}^2}}}
+{\omega _1}\left( {\frac{1}{2}G\left( 1 \right) + \sum\limits_{k =
1}^\infty {\frac{{{B_{2k}}}}{{\left( {2k} \right)!}}{G^{\left( {2k - 1}
\right)}}\left( 1 \right)} } \right) \hfill  \notag \\
&= \frac{1}{2}\ln \frac{{\omega _D^2\left( {\omega _1^2 + \omega _\mu ^2}
\right)}}{{\omega _\mu ^2{{\left( {{\omega _1} + {\omega _D}} \right)}^2}}}
+ {\omega _1}O\left( {\frac{1}{{\omega _1^2}}} \right),
\end{align}
that is, 
\begin{equation}
{R_{j\mu }} = \ln \frac{{\sqrt {\omega _1^2 + \omega _\mu ^2} }}{{{\omega _1%
} + {\omega _D}}} + {\omega _1}O\left( {\frac{1}{{\omega _1^2}}} \right).
\end{equation}
Thus, we can conclude that this estimate is relatively accurate for large and
small $\omega _1$ or more directly, $T_j$.

\section{Consistency with the Second Law of Thermodynamics}

\label{appendixC}

In fact, one can find that $\delta _{\mu }$ could be negative when $T_{1}$
and $T_{2}$ are very small. Thus one could suspect that $\omega _{\mu
}+\delta _{\mu }$ could be negative, which will lead to the violation of the
second law of thermodynamics. However, we can show that ${\omega _{\mu }}+{%
\delta _{\mu }}$ is always positive. From Eq.(\ref{32}), we get that ${%
R_{j\mu }}>-\ln ({\omega _{D}}/{\omega _{\mu }}).$ Based on \ Eqs. (\ref{P1}) (\ref{P2}) (\ref{Q}), we can obtain 
\begin{gather}
{P_{1}}>-\frac{{2{J_{1}}\left( {{\omega _{1}}}\right) }}{\pi }{\sin ^{2}}{%
\phi _{+}}\ln \frac{{{\omega _{D}}}}{{{\omega _{1}}}}=-O\left( \gamma
\right) ,\hfill \\
{P_{2}}>-\frac{{2{J_{1}}\left( {{\omega _{2}}}\right) }}{\pi }{\cos ^{2}}{%
\phi _{+}}\ln \frac{{{\omega _{D}}}}{{{\omega _{2}}}}=-O\left( \gamma
\right) ,\hfill \\
{Q_{1}}\frac{{{\omega _{D}}}}{\pi }{R_{2,1}}>-\frac{{2{J_{2}}\left( {{\omega
_{1}}}\right) }}{\pi }{\cos ^{2}}{\phi _{-}}\ln \frac{{{\omega _{D}}}}{{{%
\omega _{1}}}}=-O\left( \gamma \right) ,\hfill \\
{Q_{2}}\frac{{{\omega _{D}}}}{\pi }{R_{2,2}}>-\frac{{2{J_{2}}\left( {{\omega
_{2}}}\right) }}{\pi }{\sin ^{2}}{\phi _{-}}\ln \frac{{{\omega _{D}}}}{{{%
\omega _{2}}}}=-O\left( \gamma \right) .
\end{gather}
Therefore, one can estimate ${\omega _{\mu }}+{\delta _{\mu }}$ as 
\begin{equation}
{\omega _{\mu }}+{\delta _{\mu }}>{\omega _{\mu }}+{P_{\mu }}+\frac{{{\omega _{D}}}}{\pi }{Q_{\mu }R_{2,\mu }}>{\omega _{\mu }}-O\left( \gamma \right) >0,
\label{56}
\end{equation}
which ensures consistency with the second law of thermodynamics.


\reftitle{References}
\bibliography{book}

\isAPAStyle{%

}{}

\PublishersNote{}
\end{document}